\documentclass[journal,twoside,final]{IEEEtran}

\usepackage{algorithmic}
\usepackage{amsmath,amssymb,amsfonts,amsthm}
\usepackage{balance}
\usepackage{bm}
\usepackage{cite}
\usepackage{colortbl}
\usepackage[T1]{fontenc}
\usepackage{graphicx}
\usepackage{makecell}
\usepackage{mathrsfs}
\usepackage{mathtools}
\usepackage{multirow}
\usepackage{siunitx}
\usepackage{soul}               
\usepackage{textcomp}
\usepackage{threeparttable}
\usepackage[color=green!40]{todonotes}
\soulregister\cite7
\soulregister\footnote7
\soulregister\eqref7
\soulregister\ref7
\usepackage{xcolor}
\usepackage{url}
\usepackage{hyperref}
\newtheorem{definition}{Definition}
\newtheorem{lemma}{Lemma}
\newtheorem{theorem}{Theorem}

\newcommand{\td}{\triangledown}
\newcommand{\otd}{\overline{\triangledown}}

\graphicspath{{./}{../figures/}}

\newcommand{\myauthor}{Kyeong Soo Kim,~\IEEEmembership{Senior~Member,~IEEE}, and
  Seungyeop Kang}%
\newcommand{\mytitle}{Clock Skew Compensation Algorithm Immune to Floating-Point
  Precision Loss}%

\begin{document}

\title{\LARGE \mytitle}

\author{%
  \myauthor%
  \thanks{%
    This work was supported in part by the Postgraduate Research Scholarships
    (under Grant PGRS1912001) and the Key Programme Special Fund (under Grant
    KSF-E-25) of Xi'an Jiaotong-Liverpool University.

    K. S. Kim is with the Department of Communications and Networking, School of
    Advanced Technology, Xi'an Jiaotong-Liverpool University, Suzhou 215123,
    P. R. China (e-mail: Kyeongsoo.Kim@xjtlu.edu.cn).
    
    S. Kang is with the Department of Mechatronics and Robotics, School of
    Advanced Technology, Xi'an Jiaotong-Liverpool University, Suzhou 215123,
    P. R. China (e-mail: S.Kang18@student.xjtlu.edu.cn).%
  }%
}%


\maketitle

\begin{abstract}
  We propose a novel clock skew compensation algorithm based on Bresenham's line
  drawing algorithm. The proposed algorithm can avoid the effect of limited
  floating-point precision (e.g., 32-bit single precision) on clock skew
  compensation and thereby provide high-precision time synchronization even with
  resource-constrained sensor nodes in wireless sensor networks.
\end{abstract}


\section{Introduction}
\label{sec:introduction}
\IEEEPARstart{C}{lock} skew compensation is an essential component of time
synchronization in wireless sensor networks (WSNs), which provides a common time
frame among network nodes \cite{wu11:_clock_synch_wirel_sensor_networ}. Because
typical clock skew compensation algorithms are involved with floating-point
arithmetic, their performance on the platforms with lower computational
resources---e.g., 32-bit single-precision floating-point format on
resource-constrained WSN sensor nodes---is not up to the predicted performance
from theories or simulation experiments \cite{Kim:20-1}.

To address the issue of the limited precision floating-point arithmetic in clock
skew compensation, therefore, we propose a novel scheme based on Bresenham's
line drawing algorithm \cite{bresenham65:_algor} immune to floating-point
precision loss.

\section{Clock Models}
\label{sec:clock-models}
Without loss of generality, we confine our discussions to a network with one
head node and one sensor node in this letter, where we describe the hardware
clock $T$ of the sensor node with respect to the reference clock $t$ of the head
node using the first-order affine clock model \cite{rajan11:_joint}:
\begin{equation}
  \label{eq:hardware_clock_model}
  T(t) = \left(1+\epsilon\right)t + \theta,
\end{equation}
where $\epsilon{\in}\mathbb{R}$ and $\theta{\in}\mathbb{R}$ denote the clock
skew and offset, respectively; $(1+\epsilon){\in}\mathbb{R}_{+}$ in
\eqref{eq:hardware_clock_model} is also called \textit{clock frequency ratio} in
the literature. Because we focus on the clock skew compensation, we simplify
\eqref{eq:hardware_clock_model} by setting $\theta$ to 0 as follows:
\begin{equation}
  \label{eq:hardware_clock_model_wo_offset}
  T(t) = \left(1+\epsilon\right)t.
\end{equation}
Because our proposal in this paper compensates for the clock skew only
(equivalently clock frequency ratio), it is different from ratio-based schemes
like ratio-based time synchronization protocol (RSP)~\cite{sheu08:_ratio}
compensating both clock skew and clock offset. Note that the time
synchronization schemes based on the reverse two-way message exchange proposed
in \cite{Kim:17-1} are based on the offset-free clock model of
\eqref{eq:hardware_clock_model_wo_offset}, where the clock offset is
independently compensated for at the head node, while the sensor node only
synchronizes the frequency of its logical clock to that of the reference clock.

Compensating for the clock skew from the hardware clock $T$ in
\eqref{eq:hardware_clock_model_wo_offset}, we can obtain the logical clock
$\hat{t}$ of the sensor node---i.e., the estimation of the reference clock $t$
given the hardware clock $T$---as follows: For
$t_{i}{<}t{\leq}t_{i+1}~(i{=}0,1,{\ldots})$,
\begin{equation}
  \label{eq:logical_clock_model}
  \hat{t}\Big(T(t)\Big) = \hat{t}\Big(T(t_{i})\Big)
  + \dfrac{T(t)-T(t_{i})}{1 + \hat{\epsilon}_{i}},
\end{equation}
where $t_{i}$ is the reference time for the $i$th synchronization and
$\hat{\epsilon}_{i}$ is the estimated clock skew from the $i$th
synchronization.\footnote{The sensor node does not know the reference clock $t$
  as such but only the hardware clock corresponding to $t$---i.e.,
  $T(t)$---during the operation.}

The impact of the limited precision in floating-point arithmetic on clock skew
compensation is investigated in~\cite{Kim:20-1}: The major finding is that the
division of floating-point numbers in~\eqref{eq:logical_clock_model} incurs
substantial precision loss at typical WSN platforms with 32-bit single
precision. In case of the time synchronization scheme based on the reverse
two-way message exchange~\cite{Kim:17-1}, because the logical clock updates at
sensor nodes in~\eqref{eq:logical_clock_model} requires accurate floating-point
division and has a recursive nature, the impact of the precision loss on the
logical clock is accumulated over time.

\section{Clock Skew Compensation based on Bresenham's Algorithm}
\label{sec:clock-skew-comp-bresenhem}
The major issue in the clock skew compensation based on
\eqref{eq:logical_clock_model} is the floating-point division required for the
calculation of the second term $\frac{T(t){-}T(t_{i})}{1{+}\hat{\epsilon}_{i}}$,
i.e., the skew-compensated increment of the hardware clock since the $i$th
synchronization. In this section, we describe how to obtain the second term of
\eqref{eq:logical_clock_model} using only integer addition/subtraction and
comparison based on the Bresenham's algorithm.

Note that, though the modeling of clocks in Section~\ref{sec:clock-models} is
based on the continuous-time affine clock model, clocks in digital communication
systems are basically discrete counters \cite{etzlinger17:_times}; timestamp
values exchanged between a head and sensor nodes or recorded for events and
measurements at sensor nodes are based on the readings of discrete counters.

\begin{figure}[!tb]
  \centering%
  \includegraphics[angle=-90,width=\linewidth]{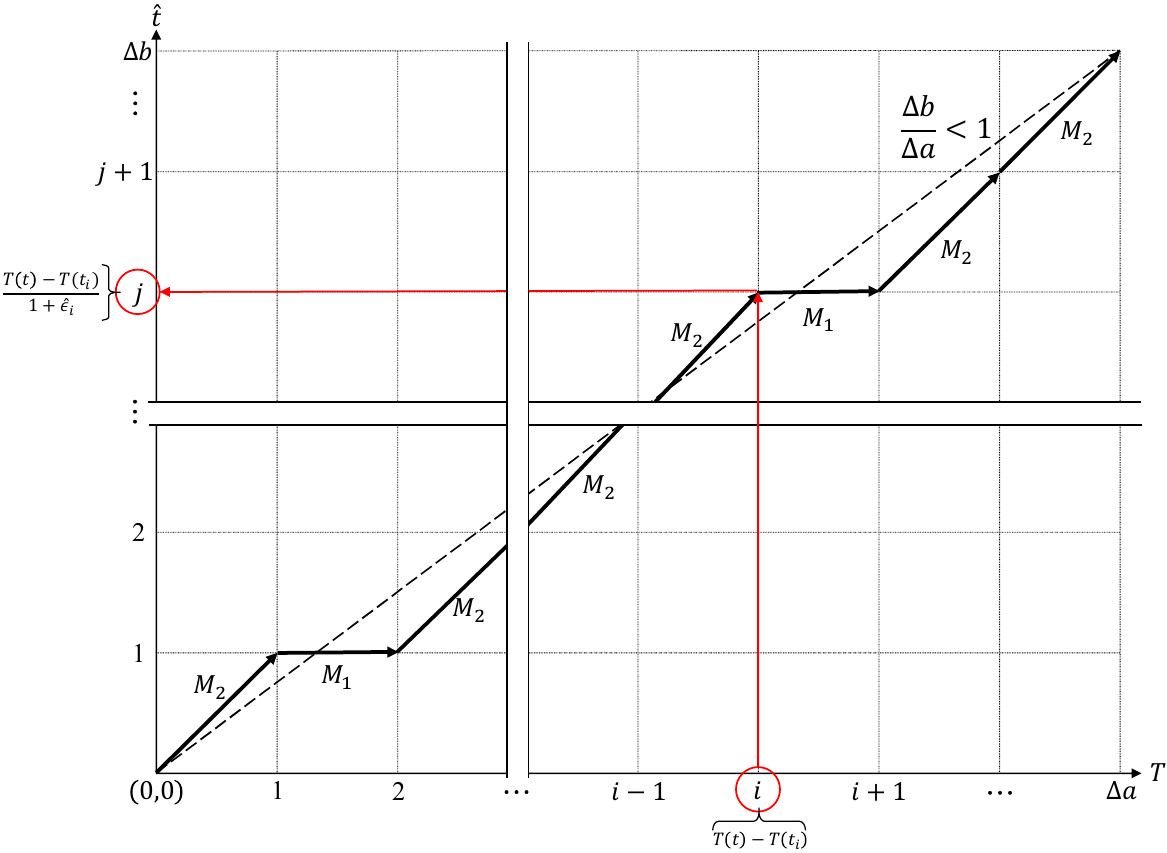}
  \caption{Clock skew compensation based on Bresenham's line drawing algorithm
    \cite{bresenham65:_algor} for the case of
    $\frac{\Delta{b}}{\Delta{a}}{<}1$.}
  \label{fig:clock_skew_compensation}
\end{figure}

Let $\frac{D}{A}$ be the inverse of a clock frequency ratio (i.e.,
$\frac{1}{1{+}\epsilon_{i}}$) estimated based on two positive integers $D$ and
$A$; $D$ and $A$ represent interdeparture and interarrival times of packets or
their cumulative sums from the previous
synchronization~\cite{Kim:13-1}.\footnote{In the following, we do not consider
  the effect of the random noise components during the estimation of the clock
  frequency ratio in the values of $D$ and $A$ to clearly assess the improvement
  made by the proposed algorithm over conventional clock skew compensation based
  on floating-point division.} Based on the Bresenham's
algorithm~\cite{bresenham65:_algor}, we can obtain the skew-compensated
increment of the hardware clock as follows: For $\frac{D}{A}{<}1$, we set
$\Delta{a}$ and $\Delta{b}$ to $A$ and $D$, respectively, and calculate
$\td_{i}$ using the following recursive relation:\footnote{The case of
  $\frac{D}{A}{>}1$ will be handled in
  Theorem~\ref{thm:clock_skew_compensation}, and the index starts from 0 to make
  mapping easier between $\td_{i}$ and the $x$ coordinate of a point (e.g.,
  $\td_{0}$ is for $(0,0)$).}
\begin{align}
  \td_{0} & = 2\Delta b - \Delta a, \label{eq:bresenham_algo_1}\\
  \td_{i+1} & = \begin{cases}
    \td_{i} + 2\Delta b - 2\Delta a & \text{if $\td_{i} \geq 0$}, \\
    \td_{i} + 2\Delta b & \text{otherwise}.
  \end{cases} \label{eq:bresenham_algo_2}
\end{align}
Then, from the origin and on, we determine each movement based on
$\td_{i}$:
\begin{equation}
  \label{eq:movement}
  \begin{cases}
    M_{1} & \text{if $\td_{i} < 0$}, \\
    M_{2} & \text{otherwise},
  \end{cases}
\end{equation}
where $M_{1}$ is a horizontal movement and $M_{2}$ is a diagonal movement shown
in Fig.~\ref{fig:clock_skew_compensation}.\footnote{The coordinates are shifted
  by a pair of the initial values of hardware and logical clocks, e.g.,
  $\left(T(t_{i}),\hat{t}\left(T(t_{i})\right)\right)$ in
  \eqref{eq:logical_clock_model}.}
Lemma~\ref{lem:td_boundedness} shows that $\td_{i}$ is bounded.
\begin{lemma}
  \label{lem:td_boundedness}
  $\td_{i}$ satisfies the following inequality:
  \begin{equation}
    \label{eq:td_boundedness}
    \left|\td_{i}\right| < 2\Delta a.
  \end{equation}
\end{lemma}
\begin{IEEEproof}
  See Appendix~\ref{sec:proof-lemma-1}.
\end{IEEEproof}

Unlike the line drawing, the skew compensation in our case does not need all the
intermediate points between the origin and the point under consideration; we
need only the $y$ coordinate of a point given its $x$ coordinate (e.g., $j$
given $i$ in Fig.~\ref{fig:clock_skew_compensation}), where $x$ coordinate is
the increment of the hardware clock and $y$ coordinate is its skew-compensated
increment for the logical clock $\hat{t}(T)$. Therefore, we cannot use
Bresenham's original algorithm as it is, especially when $D$ and $A$ are large
and there are sparse events/measurements.

\begin{figure}[!tb]
  \centering%
  \includegraphics[angle=-90,width=.7\linewidth]{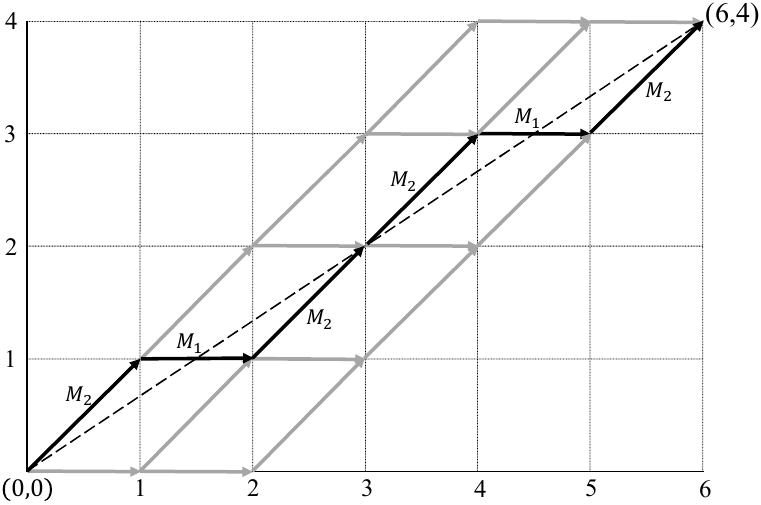}
  \caption{Example of all possible paths from the origin to a given destination
    point based on $M_{1}$ and $M_{2}$ movements.}
  \label{fig:possible_paths}
\end{figure}

To extend Bresenham's algorithm for skew compensation, let's consider the
example shown in Fig.~\ref{fig:possible_paths}, where the only valid path from
$(0,0)$ to $(6,4)$ according to Bresenham's algorithm is indicated by black
arrows while alternative paths are by gray arrows.
Though the alternative paths are not useful for line drawing, they can reach the
same destination and thereby provide the correct $y$ coordinate. In fact,
Lemma~\ref{lem:td_boundedness} shows that $\td_{i}$ in Bresenham's algorithm
depends only on the number of $M_{2}$, not the exact sequence of $M_{1}$ and
$M_{2}$, during the total $i$ movements. Therefore, we extend $\td_{i}$ of
Bresenham's algorithm to the points on all possible paths from the origin to a
given destination of $(\Delta{a},\Delta{b})$ based on the two movements of
$M_{1}$ and $M_{2}$ and define $\otd_{i}(j)$ as follows:
\begin{definition}
  \label{def:vp_set}
  Given a destination point $(\Delta{a},\Delta{b})$, we define valid point set
  $\mathscr{V}(\Delta{a},\Delta{b})$ as a set of the points on a valid path from
  $(0,0)$ to $(\Delta{a},\Delta{b})$ according to Bresenham's algorithm.
\end{definition}
\begin{definition}
  \label{def:br_set}
  Given a point $(i,j){\in}\mathscr{V}(\Delta{a},\Delta{b})$, we define
  backward-reachable set $\mathscr{B}(i,j)$ as a set of the points that can
  reach $(i,j)$ by any combination of the movements $M_{1}$ and $M_{2}$ as
  follows:
  \begin{multline}
    \label{eq:br_set}
    \mathscr{B}(i,j) \triangleq \\
    \left\{(k,l)|0{\leq}k{<}i,\max(0,k{-}\Delta{a}{+}\Delta{b}){\leq}l{\leq}\min(k,j)\right\}.
  \end{multline}
\end{definition}
\begin{definition}
  \label{def:extended_td}
  For a point $(i,j){\in}\mathscr{B}(\Delta{a},\Delta{b})$, we define $\otd_{i}(j)$ as follows:
  \begin{equation}
    \label{eq:non-recursive_td}
    \otd_{i}(j) \triangleq 2(i\Delta b - j\Delta{a}).
  \end{equation}
\end{definition}

Note that
$\otd_{i}(j){=}\td_{i},{\forall}(i,j){\in}\mathscr{V}(\Delta{a},\Delta{b})$.
Lemma~\ref{lem:td_reachability} shows the property of $\otd_{i}(j)$ essential to
the extension of Bresenham's algorithm to clock skew compensation:
\begin{lemma}
  \label{lem:td_reachability}
  A point $(i,j){\in}\mathscr{V}(\Delta{a},\Delta{b})$ is reachable from any point
  $(k,l){\in}\mathscr{B}(i,j)$ if we apply
  \eqref{eq:bresenham_algo_1}--\eqref{eq:movement} using $\otd_{\cdot}(\cdot)$
  instead of $\td_{\cdot}$.
\end{lemma}
\begin{IEEEproof}
  See Appendix~\ref{sec:proof-lemma-2}.
\end{IEEEproof}

Now we can prove the main theorem for the clock skew compensation based on the
extension of Bresenham's algorithm, which can eliminate the effect of precision
loss on floating-point arithmetic:
\begin{theorem}
  \label{thm:clock_skew_compensation}
  Given the hardware clock $i$, we can obtain its skew-compensated clock $j$ as follows:

  \smallskip
  \noindent
  \textit{Case 1.} $\frac{D}{A}{<}1$: The skew-compensated clock $j$ satisfies
  \begin{equation}
    \label{eq:new_skew_compensation-1}
    i\frac{D}{A} - 1 < j < i\frac{D}{A} + 1.
  \end{equation}
  Unless $i\frac{D}{A}$ is an integer, there are two values satisfying
  \eqref{eq:new_skew_compensation-1}. Due to the effect of limited
  floating-point precision, however, we cannot know the exact value of
  $i\frac{D}{A}$. In this regard, \eqref{eq:new_skew_compensation-1} can be
  extended to include the effect of the precision loss:
  \begin{equation}
    \label{eq:new_skew_compensation-2}
    i\frac{D}{A} - 1 - \varepsilon < j < i\frac{D}{A} + 1 + \varepsilon,
  \end{equation}
  where $\varepsilon({>}0)$ is the error due to the precision loss. Let
  $k,{\ldots},k{+}l$ be the candidate values of $j$ satisfying
  \eqref{eq:new_skew_compensation-2}. We determine $j$ by starting from the
  point $(i{-}l,k)$ and applying Bresenham's algorithm with $\otd_{i-l}(k)$ and
  on; $j$ is determined by the $y$ coordinate of the valid point whose $x$
  coordinate is $i$.

  \smallskip
  \noindent
  \textit{Case 2.} $\frac{D}{A}{>}1$: In this case, we can decompose the
  skew-compensated clock $j$ into two components as follows:
  \begin{equation}
    \label{eq:decomposition}
    j = i\frac{D}{A} = i + i\frac{D - A}{A}.
  \end{equation}
  Now that $\frac{D{-}A}{A}{<}1$, we can apply the same procedure of Case~1 to
  the second component in \eqref{eq:decomposition} by setting $\Delta{a}$ and
  $\Delta{b}$ to $A$ and $D{-}A$, respectively. Let $\bar{j}$ be the result from
  the procedure. The skew-compensated clock $j$ is given by $i{+}\bar{j}$ as per
  \eqref{eq:decomposition}.
\end{theorem}
\begin{IEEEproof}
  See Appendix~\ref{sec:proof-theorem-1}.
\end{IEEEproof}

\begin{figure}[!tb]
  \centering%
  \includegraphics[angle=-90,width=.6\linewidth]{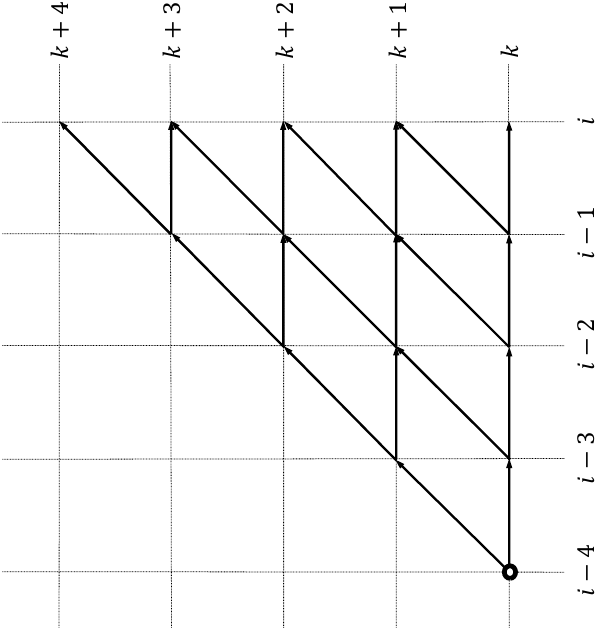}
  \caption{Example of common starting points reaching all possible candidate
    points for the case of $\frac{\Delta{b}}{\Delta{a}}{<}1$.}
  \label{fig:starting_points}
\end{figure}

In summary, the clock skew compensation algorithm described in
Theorem~\ref{thm:clock_skew_compensation} allows us to bound the correct
skew-compensated clock given a hardware clock by
\eqref{eq:new_skew_compensation-2} and search for it from a nearby starting
point---instead of the origin---using the extension of Bresenham's algorithm
based on $\otd_{\cdot}(\cdot)$. Fig.~\ref{fig:starting_points} illustrates an
example of the starting point discussed in
Theorem~\ref{thm:clock_skew_compensation}.

Note that the case of $\frac{D}{A}{>}1$ is handled by setting $\Delta{a}$ and
$\Delta{b}$ to $D$ and $A$ in Bresenham's original algorithm, where we need to
find the skew-compensated clock $j$ on the $x$-axis given the hardware clock $i$
on the $y$-axis. The major issue is that there could be multiple points on the
$x$-axis given a point in the $y$-axis due to the movement of $M_{1}$ (e.g.,
$(1,1)$ and $(2,1)$ in Fig.~\ref{fig:clock_skew_compensation}). The algorithm
described in Theorem~\ref{thm:clock_skew_compensation} can avoid such an issue
resulting from the change of the axes by unifying both cases with the same
procedure for $\frac{D}{A}{<}1$.

\subsection{Numerical Examples}
\label{sec:numerical-examples}
We apply the proposed algorithm to the representative cases whose results are
summarized in Table~\ref{tab:csc_results}.
\begin{table}[!tb]
  \centering
  \begin{threeparttable}
    \caption{Results of Clock Skew Compensation.}
    \label{tab:csc_results}
    {%
      \renewcommand{\arraystretch}{1.3}
      \begin{tabular}{c|r|r|r|r|r|r}
        \hline
        \multirow{2}{*}{Algorithm} & \multicolumn{1}{c|}{\multirow{2}{*}{$i$}}
        & \multicolumn{3}{c|}{Compensation error\tnote{*}} & \multicolumn{2}{c}{\# of iterations\tnote{\dag}} \\
        \cline{3-7}
                                   & & \multicolumn{1}{c|}{Min.} & \multicolumn{1}{c|}{Max.} & \multicolumn{1}{c|}{Avg.} & \multicolumn{1}{c|}{Min.} & \multicolumn{1}{c}{Max.} \\
        \hline
                                   & \num{1e6} & \num{0} & \num{0} & \num{0} &
                                                                               --
                                                                 & --
        \\
        Single & \num{1e7} & \num{0} & \num{0} & \num{0} & -- & -- \\
        precision\tnote{\ddag} & \num{1e8} & \num{-4} & \num{1} & \num{-2.0004}
                                     & --  & -- \\
                                   & \num{1e9} & \num{-19} & \num{44} &
                                                                        \num{1.2382e1}
                                     & -- & -- \\
        \hline
        \multirow{4}{*}{Proposed} & \num{1e6} & \num{-1} & \num{0} &
                                                                     \num{-4.9820e-01}
                                     & \num{2} & \num{2} \\
                                   & \num{1e7} & \num{-1} & \num{0} &
                                                                      \num{-4.9685e-01}
                                     & \num{3} & \num{4} \\
                                   & \num{1e8} & \num{-1} & \num{0} &
                                                                      \num{-5.0263e-01}
                                     & \num{21} & \num{22} \\
                                   & \num{1e9} & \num{-1} & \num{0} &
                                                                      \num{-4.9708e-01}
                                     & \num{201} & \num{202} \\
        \hline
      \end{tabular}
    }%
    \begin{tablenotes}
    \item[*] With respect to ${\lfloor}i\frac{D}{A}{\rfloor}$ based on double
      precision.
    \item[\dag] $l$ in Theorem~\ref{thm:clock_skew_compensation}.
    \item[\ddag] ${\lfloor}i\frac{D}{A}{\rfloor}$ based on single precision.
    \end{tablenotes}
  \end{threeparttable}
\end{table}
We fix $D$ to $1,000,000$ and generate one million samples of $A$ whose clock
skews are uniformly distributed in the range of
$[{-}100\,\text{ppm},100\,\text{ppm}]$~\cite{instrument02:_selec_texas_instr_usb}. The
value of $\varepsilon$ is set to $10^{-7}i$ based on the analysis in
\cite[Section~2]{Kim:20-1}; the minimum and the maximum values of $i$ (i.e., the
hardware clock) in Table~\ref{tab:csc_results} correspond to \SI{1}{\s} and
\SI{1000}{\s}, respectively, at a sensor node running TinyOS whose
synchronization limit is \SI{1}{\us} \cite{tinyos}. For the clock skew
compensation by single and double-precision floating-point arithmetic, we round
down the results to obtain integer values.

The results in Table~\ref{tab:csc_results} show that the clock skew compensation
errors of the proposed algorithm with respect to the results based on
double-precision floating-point arithmetic are bounded by $[{-}1,0]$, while
those of the single-precision algorithm begins to increase when the hardware
clock is \num{1e8}. When we change rounding down to rounding off and rounding up
for the results from double-precision and single-precision algorithms, we still
obtain similar results where the compensation errors of the proposed algorithm
are always bounded by $[{-}1,1]$.

Note that $i{=}\num{1e9}$---corresponding to \SI{1000}{\s}---is unlikely in a
practical scenario because the maximum value of $i$ is limited by the
synchronization interval, especially considering the cheap quartz crystal
oscillator. Excluding the case of $i{=}\num{1e9}$, therefore, we observe that
the number of iterations for a typical value of $i$ is reasonable and that the
computational complexity of the proposed algorithm is not a major issue. During
the experiments, we also found that the upper and lower bounds in
\eqref{eq:new_skew_compensation-2} based on $\varepsilon{=}10^{-7}i$ are very
loose and that there is room for tighter bounds and thereby reducing the number
of iterations, which would require further investigation on the tradeoff between
computational complexity and precision loss in the proposed algorithm.

\section{Conclusions}
\label{sec:conclusions}
In this letter, we have proposed a novel clock skew compensation algorithm
immune to floating-point precision loss, where we extend Bresenham's algorithm
to the points on all possible paths from the origin to the destination based on
the movements of $M_{1}$ and $M_{2}$ and unify the two cases of
$\frac{D}{A}{<}1$ and $\frac{D}{A}{>}1$ with a common procedure. Through
numerical examples, we have also demonstrated that the proposed algorithm can
compensate for clock skew without being affected by precision loss unlike the
existing schemes based on floating-point divisions.

\appendices
\section{Proof of Lemma~1}
\label{sec:proof-lemma-1}
Because $0{<}\frac{\Delta{b}}{\Delta{a}}{<}1$,
\begin{align}
  \label{eq:boundedness_td1}
  \begin{split}
    0 & < \Delta b < \Delta a, \\
    0 & < 2\Delta b < 2\Delta a, \\
    -\Delta a & < 2\Delta b - \Delta a < \Delta a, \\
    \therefore -2\Delta a & < \td_{0} < 2\Delta a.
  \end{split}
\end{align}

Let's assume ${-}2\Delta{a}{<}\td_{i}{<}2\Delta{a}$ for $i{\geq}0$.

\smallskip
\noindent
\textit{Case 1.} $\td_{i}{\geq}0$:
\begin{align}
  \label{eq:boundedness_tdi-1}
  \begin{split}
    0 & \leq \td_{i} < 2\Delta a, \\
    2\Delta b - 2\Delta a & \leq \td_{i} + 2\Delta b - 2\Delta a < 2\Delta b.
  \end{split}
\end{align}
From \eqref{eq:boundedness_td1}, we have
${-}2\Delta{a}{<}2\Delta{b}{-}2\Delta{a}$ and $2\Delta{b}{<}2\Delta{a}$. So we
obtain
\begin{align}
  \label{eq:boundedness_tdi-2}
  \begin{split}
    -2\Delta a & < \td_{i} + 2\Delta b - 2\Delta a < 2\Delta a, \\
    \therefore -2\Delta a & < \td_{i+1} < 2\Delta a.
  \end{split}
\end{align}

\smallskip
\noindent
\textit{Case 2.} $\td_{i}{<}0$:
\begin{align}
  \label{eq:boundedness_tdi-3}
  \begin{split}
    -2\Delta a & < \td_{i} < 0, \\
    2\Delta b - 2\Delta a & < \td_{i} + 2\Delta b < 2\Delta b. \\
  \end{split}
\end{align}
Again, from \eqref{eq:boundedness_td1}, we have
${-}2\Delta{a}{<}2\Delta{b}{-}2\Delta{a}$ and $2D<2A$. So we obtain
\begin{align}
  \label{eq:boundedness_tdi-4}
  \begin{split}
    -2\Delta a & < \td_{i} + 2\Delta b < 2\Delta a, \\
    \therefore -2\Delta a & < \td_{i+1} < 2\Delta a.
  \end{split}
\end{align}
\eqref{eq:boundedness_td1}--\eqref{eq:boundedness_tdi-4} completes the proof by
mathematical induction. \hfill\IEEEQED

\section{Proof of Lemma~\ref{lem:td_reachability}}
\label{sec:proof-lemma-2}
We first show that $(i,j){\in}\mathscr{V}(\Delta{a},\Delta{b})$ is reachable from
$(i{-}1,l){\in}\mathscr{B}(i,j)$, i.e., the points one step backward from it.
Considering the movements of $M_{1}$ and $M_{2}$, we could have at most two such
points, i.e., $(i{-}1,j{-}1)$ and $(i{-}1,j)$.\footnote{It is possible that
  $(i{-}1,j){\notin}\mathscr{B}$ (e.g., $(1,1)$ in
  Fig.~\ref{fig:possible_paths}).}

\smallskip
\noindent
\textit{Case 1.} $(i{-}1,j{-}1){\in}\mathscr{V}(\Delta{a},\Delta{b})$:
$\td_{i-1}$ should be equal to or greater than 0 so that we take $M_{2}$ and
move to $(i,j)$. Now we have two subcases:

\smallskip
\noindent
\textit{Case 1.1.} Start from $(i{-}1,j{-}1)$: $(i,j)$ can be reachable using
$\otd_{i-1}(j{-}1)$ because $\otd_{i-1}(j{-}1){=}\td_{i-1}$.

\smallskip
\noindent
\textit{Case 1.2.} Start from $(i{-}1,j)$:
$\otd_{i-1}(j){=}\otd_{i-1}(j{-}1){-}2\Delta{a}$, so we have
\begin{align}
  \label{eq:proof-lemma-2-case1-2}
  \begin{split}
    -2\Delta a & < \otd_{i-1}(j-1) < 2\Delta a ~ (\because \otd_{i-1}(j{-}1){=}\td_{i-1}), \\
    -4\Delta a & < \otd_{i-1}(j-1) - 2\Delta a < 0, \\
    -4\Delta a & < \otd_{i-1}(j) < 0.
  \end{split}
\end{align}
Because $\otd_{i-1}(j){<}0$, we take $M_{1}$ and move to $(i,j)$.

\smallskip
\noindent
\textit{Case 2.} $(i{-}1,j){\in}\mathscr{V}(\Delta{a},\Delta{b})$: $\td_{i-1}$ should be less
than 0 so that we take $M_{1}$ and move to $(i,j)$. Again, we have two subcases:

\smallskip
\noindent
\textit{Case 2.1.} Start from $(i{-}1,j{-}1)$:
$\otd_{i-1}(j{-}1){=}\otd_{i-1}(j){+}2\Delta{a}$, so we have
\begin{align}
  \label{eq:1}
  \begin{split}
    -2\Delta a & < \otd_{i-1}(j) < 2\Delta a ~ (\because \otd_{i-1}(j){=}\td_{i-1}), \\
    0 & < \otd_{i-1}(j) + 2\Delta a < 4\Delta a, \\
    0 & < \otd_{i-1}(j{-}1) < 4\Delta a.
  \end{split}
\end{align}
Because $\otd_{i-1}(j{-}1){>}0$, we take $M_{2}$ and move to $(i,j)$.
  
\smallskip
\noindent
\textit{Case 2.2.} Start from $(i{-}1,j)$: $(i,j)$ can be reachable
using $\otd_{i-1}(j)$ because $\otd_{i-1}(j){=}\td_{i-1}$.

\medskip

Now we assume that $(i,j){\in}\mathscr{V}(\Delta{a},\Delta{b})$ is reachable
from $(i{-}k,l){\in}\mathscr{B}(i,j)$ for $k{\geq}1$, i.e., the points $k$ step
backward from it. Because in general we can move to
$(i-k,l){\in}\mathscr{B}(i,j)$ from the points $k{+}1$ step back from $(i,j)$ by
taking either $M_{1}$ or $M_{2}$, we consider only two special boundary cases of
$(\Delta{a}{-}\Delta{b},0)$ and $(\Delta{b},\Delta{b})$\footnote{See the points
  $(2,0)$ and $(4,4)$ in Fig.~\ref{fig:possible_paths} as an example.} to check
whether the movement from those points still belongs to $\mathscr{B}(i,j)$.

\smallskip
\noindent
\textit{Case 1.} $(i{-}k{-}1,l){=}(\Delta{a}{-}\Delta{b},0)$: We take $M_{2}$
and the next point belongs to $\mathscr{B}(i,j)$ because
$\otd_{\Delta{a}-\Delta{b}}(0){=}2\Delta{b}(\Delta{a}{-}\Delta{b}){>}0$.

\smallskip
\noindent
\textit{Case 2.} $(i{-}k{-}1,l){=}(\Delta{b},\Delta{b})$: We take $M_{1}$ and
the next point belongs to $\mathscr{B}(i,j)$ because
$\otd_{\Delta{b}}(\Delta{b}){=}2\Delta{b}(\Delta{b}{-}\Delta{a}){<}0$.

\medskip
This completes the proof by mathematical induction. \hfill\IEEEQED

\section{Proof of Theorem~\ref{thm:clock_skew_compensation}}
\label{sec:proof-theorem-1}
Here we prove only Case~1 because Case~2 follows from Case~1 and is just its
application.

\textit{Case 1.} $\frac{D}{A}{<}1$: If
$(i,j){\in}\mathscr{V}(\Delta{a},\Delta{b})$, we obtain the following from
Lemma~\ref{lem:td_boundedness} and Definition~\ref{def:extended_td}:
\begin{align}
  \label{eq:2}
  \begin{split}
    -2\Delta a & < \otd_{i}(j) < 2\Delta a, \\
    -2\Delta a & < 2(i\Delta b - j\Delta a) < 2\Delta a, \\
    -2A & < 2(iD - jA) < 2A ~ (\because \Delta{a}{=}A, \Delta{b}{=}D).\\
  \end{split}
\end{align}
Hence \eqref{eq:new_skew_compensation-1}. If $i\frac{D}{A}$ is not an integer,
it is clear from \eqref{eq:new_skew_compensation-1} that there are two values
satisfying the inequality. If we also include the effect of the precision loss,
there could be even more values satisfying the extended inequality of
\eqref{eq:new_skew_compensation-2}.

Let $k,{\ldots},k{+}l$ be those values satisfying
\eqref{eq:new_skew_compensation-2}. Because $(i{-}l,k)$ belongs to
$\bigcap_{m=0}^{l}\mathscr{B}(i,k{+}m)$ as per \eqref{eq:br_set}, we can reach
from $(i{-}l,k)$ to a valid point---i.e., one of $(i,k),{\ldots},(i,k{+}l)$---by
applying Bresenham's algorithm with $\otd_{i-l}(k)$ by
Lemma~\ref{lem:td_reachability}. \hfill\IEEEQED

\balance 


\end{document}